%% file: main.tex
\documentclass[sigconf, screen]{acmart}

\usepackage[normalem]{ulem}
\usepackage{enumitem}

\usepackage{microtype}

\usepackage{tabularx}
\usepackage{array}
\usepackage{listings}
\usepackage{listingsutf8}
\lstdefinestyle{promptblock}{
  basicstyle=\ttfamily\footnotesize,
  breaklines=true,
  breakatwhitespace=false,
  breakautoindent=false,
  breakindent=0pt,
  columns=fullflexible,
  keepspaces=true,
  showstringspaces=false,
  upquote=true,
  frame=none,
  xleftmargin=0pt, xrightmargin=0pt
}
\usepackage[utf8]{inputenc}

\useunder{\uline}{\ul}{}

\def\eg{\emph{e.g., }} 
\def\ie{\emph{i.e., }} 

\definecolor{DO}{HTML}{ffe3c8} 
\definecolor{DO1}{HTML}{ffe3c8} 
\definecolor{DO2}{HTML}{c2cae4} 
\definecolor{DO3}{HTML}{fecbe5} 
\definecolor{DO4}{HTML}{c9ebe7} 
\definecolor{DO5}{HTML}{f6f6cf} 

\newcommand{\sally}[1]{ \textcolor{black}{#1}}

\newcommand{\pquotes}[1]{\textcolor[gray]{0.3}{\textit{#1}}}
\newcommand{\prompts}[1]{\textcolor[gray]{0.3}{\textit{#1}}}

\makeatletter
\newcommand\footnoteref[1]{\protected@xdef\@thefnmark{\ref{#1}}\@footnotemark}
\makeatother

\AtBeginDocument{%
  }

\setcopyright{cc}
\setcctype{by}
\acmDOI{10.1145/3757279.3785640}
\acmYear{2026}
\copyrightyear{2026}
\acmISBN{979-8-4007-2128-1/2026/03}
\acmConference[HRI '26]{Proceedings of the 21st ACM/IEEE International Conference on Human-Robot Interaction}{March 16--19, 2026}{Edinburgh, Scotland, UK}
\acmBooktitle{Proceedings of the 21st ACM/IEEE International Conference on Human-Robot Interaction (HRI '26), March 16--19, 2026, Edinburgh, Scotland, UK}
\received{2025-09-30}
\received[accepted]{2025-12-01}

\begin{document}

\title[Reframing Conversational Design in HRI]{Reframing Conversational Design in HRI: \\Deliberate Design with AI Scaffolds}

\author{Shiye Cao}
\affiliation{%
  \institution{Johns Hopkins University}
  \city{Baltimore}
  \country{USA}
}

\author{Jiwon Moon}
\affiliation{%
  \institution{University of Chicago}
  \city{Chicago}
  \country{USA}
}

\author{Yifan Xu}
\affiliation{%
  \institution{Johns Hopkins University}
  \city{Baltimore}
  \country{USA}
}

\author{Anqi Liu}
\affiliation{%
  \institution{Johns Hopkins University}
  \city{Baltimore}
  \country{USA}
}

\author{Chien-Ming Huang}
\affiliation{%
  \institution{Johns Hopkins University}
  \city{Baltimore}
  \country{USA}
}

\begin{abstract}
Large language models (LLMs) have enabled conversational robots to move beyond constrained dialogue toward free-form interaction. However, without context-specific adaptation, generic LLM outputs can be ineffective or inappropriate. This adaptation is often attempted through prompt engineering, which is non-intuitive and tedious. Moreover, predominant design practice in HRI relies on impression-based, trial-and-error refinement without structured methods or tools, making the process inefficient and inconsistent. To address this, we present the AI-Aided Conversation Engine (ACE), a system that supports the deliberate design of human-robot conversations. ACE contributes three key innovations: 1) an LLM-powered voice agent that scaffolds initial prompt creation to overcome the ``blank page problem,'' 2) an annotation interface that enables the collection of granular and grounded feedback on conversational transcripts, and 3) using LLMs to translate user feedback into prompt refinements. We evaluated ACE through two user studies, examining both designs' experience and end users' interactions with robots designed using ACE. Results show that ACE facilitates the creation of robot behavior prompts with greater clarity and specificity, and that the prompts generated with ACE lead to higher-quality human-robot conversational interactions. 
\end{abstract}





\begin{teaserfigure}
    \centering
    \includegraphics[width=\textwidth]{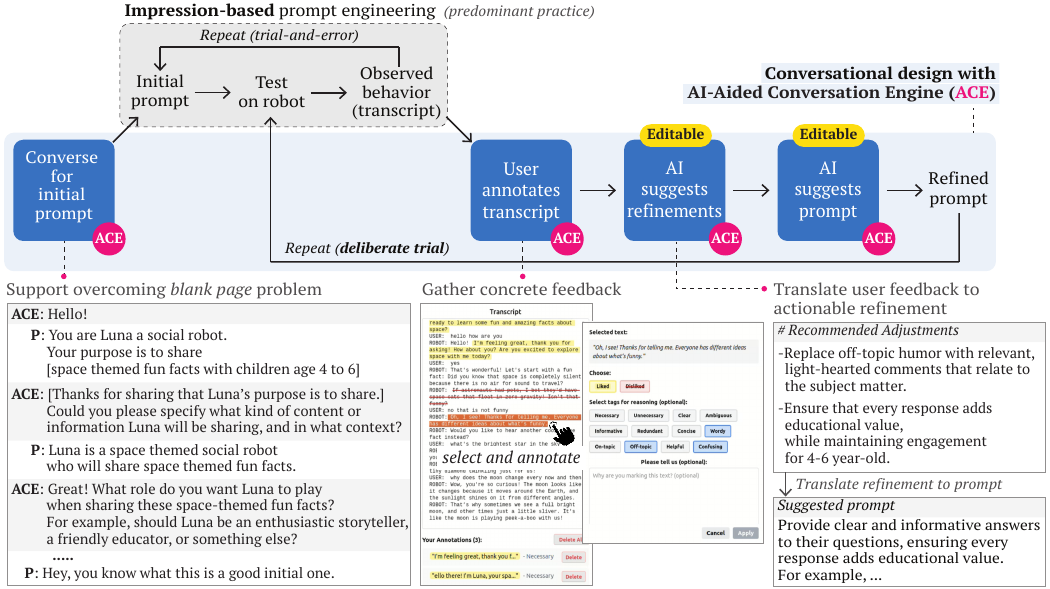}
    \caption{\sally{We augment the impression-based predominant design pipeline by 1) using an LLM-powered voice agent to scaffold initial prompt creation, 2) enabling granular feedback via transcript annotations, and 3) leveraging LLMs to translate feedback into actionable refinements for more deliberate human-robot conversation design. Bracketed text indicates overlapping speech.}}
    \label{fig:teaser}
    \Description{The figure shows two pipelines for human-robot conversation design. The first pipeline demonstrates the predominant practice with is impression-based prompt engineering. Users start by providing an initial prompt, then test the robot, and then observe behavior using the transcript. They repeat this process with trial and error. The other pipeline shows conversational design with AI-Aided Conversation Engine (ACE). ACE augments initial prompt drafting by having a conversation with user for initial prompt to overcome blank page problem. Sample conversation: ACE: Hello!, P: You are Luna a social robot. Your purpose is to share [space themed fun facts with children age 4 to 6] ACE: [Thanks for sharing that Luna’s purpose is to share.] Could you please specify what kind of content or information Luna will be sharing, and in what context? P: Luna is a space themed social robot who will share space themed fun facts. ACE: Great! What role do you want Luna to play when sharing these space-themed fun facts? For example, should Luna be an enthusiastic storyteller, a friendly educator, or something else? ..... P: Hey, you know what this is a good initial one. Text in brackets indicates overlapping speech. ACE also augments the observed behavior after testing. ACE gather concrete feedback from users through select and annotate and provides AI suggested refinements based on the feedback. Example: recommended adjustments; replace off-topic humor with relevant, light-hearted comments that relate to the subject matter; ensure that every response adds education value, while maintaining engagement for 4-6 year-old. Then, AI suggests prompt changes based on the suggested refinements, translating user feedback to actionable refinements. Example: Suggested prompt; provide clear and informative answers to their questions, ensuring every response adds education value. For example, .... Lastly, once user create a refined prompt and repeat testing the robot. This process is more deliberate design process. }
\end{teaserfigure}

\thanks{\textbf{Email}: scao14@jhu.edu. \textbf{AI Statement}. This paper was proofread using a language model; authors verified content accurately reflects their original intent. \textbf{CRediT authorship}: Conceptualization \& Method \& Software (SC, JM, YX, CH); Investigation \& Analysis (SC, JM, YX); \& Writing \& Review (all); Funding \& Supervision (CH)}

\maketitle

\input{introduction}

\input{background}

\input{system}

\input{study1}

\input{study2}

\input{discussion}

\begin{acks}
This work was in part supported by the National Science Foundation awards 2141335 and 2143704. 
\end{acks}

\newpage
\balance
\bibliographystyle{ACM-Reference-Format}
\bibliography{references}

\newpage


\end{document}

%% file: introduction.tex
\section{Introduction}
As robots are envisioned to take on a variety of roles in people's everyday life, they must be able to communicate effectively with diverse audiences across varied contexts. People naturally adapt their communication (\ie tone, word choice, formality) based on context and audience. Robots, however, can only achieve this adaptability through deliberate design. The integration of large language models (LLMs) has revolutionized robots' conversational capabilities, enabling them to engage in flexible, free-form dialogue on generic topics ``out of the box.'' However, without context-specific adaptation, generic LLM outputs can be ineffective, misaligned with task goals, or even inappropriate \cite{wang2024ain, allgeuer2024robots, kim2024understanding}. Prompt engineering offers a way to steer LLM outputs toward desired behaviors, but crafting effective prompts is challenging.

Prompt engineering is often referred to as an art, as it requires intuition, creativity, and iteration \cite{oppenlaender2025prompting}. Part of the challenge stems from the black-box nature of LLMs. Their internal mechanisms are opaque, and even small changes in the prompts can lead to large and unpredictable shifts in the output. While some guidelines exist for crafting more effective prompts, predominant practice in conversation design is based on impression-based trial-and-error, which can lead to inefficiency and decreased designer confidence due to the lack of concrete evidence to support and ground their design decisions \cite{choi2021protochat}. Hence, in this work, we re-think the design of human-robot conversations in the era of generative AI. 


\sally{We follow a five-stage design process to identify design objectives and iteratively develop an AI-Aided Conversation Engine (ACE) to support deliberate design of human-robot conversations (see Fig.~\ref{fig:teaser}). We reviewed literature and interviewed HRI researchers to understand current practices and challenges, then iteratively designed, developed, and tested system prototypes to refine design goals and the system. Our system has the following novel features:} 
\begin{itemize}[leftmargin=*, itemsep=0pt, topsep=0pt, parsep=0pt]
    \item \sally{A voice-based LLM agent to scaffold initial robot behavior prompts.}
    \item \sally{Enable collection of granular feedback on conversation transcripts to support reflection, rather than relying on impressions.} 
    \item \sally{Use of LLMs to facilitate translation of the collected feedback into prompt refinements.} 
\end{itemize}

In addition, our work has the following contributions: 
\begin{itemize}[leftmargin=*, itemsep=0pt, topsep=0pt, parsep=0pt]
    \item The design, development, and open-sourcing of ACE \footnotemark. 
    \item Empirical evidence showing that ACE can support designers in creating better robot behavior prompts. 
    \item Empirical evidence demonstrating that prompts created with ACE can lead to better human-robot conversation. 
\end{itemize}

\footnotetext{Code and additional study and implementation details available at \url{https://github.com/intuitivecomputing/hri-conversation-design}.}


%% file: background.tex
\section{Related Work}

\subsection{Human-Robot Conversation Design}
\label{sec:human-robot-conversation-design}

Traditional conversational agents relied primarily on hand-crafted scripts (\eg \cite{scassellati2018improving}) and predefined pattern–response templates (\eg\cite{csapo2012multimodal}) to respond to users. Conversation designers often collect human conversation data and feedback through wizard-of-oz studies \cite{prasad2019dara, kocielnik2018designing}, workshops \cite{kocielnik2018reflection}, and crowdsourcing \cite{choi2021protochat} to extract design insights, which is labor-intensive and requires participant recruitment. Other works created conversation templates by analyzing text or dialogue from the internet or application logs (\eg \cite{hu2018touch, zhou2020design, csapo2012multimodal, li2018kite}). Although such data are available at scale, they are difficult to apply directly to human-agent conversation design \cite{choi2021protochat}. 

LLMs have shifted conversational design from enumerating utterances to steering generative models. Pretrained on broad data, LLMs can generate responses to open-domain dialogue, enabling robots to handle flexible, free-form interactions across domains such as education, healthcare, and customer service \cite{irfan2024recommendations, ZHANG2023100131, spitale2025vita, Zhang_2023}. However, without careful design of domain adaptation, LLM-powered conversations can be misaligned with HRI needs, exhibiting superficiality, repetition, or factual hallucinations \cite{irfan2025between}. Studies show that poor prompt engineering can lead robots to struggle with maintaining engagement \cite{wang2024ain}, lack cognitive skills \cite{allgeuer2024robots}, appropriate social behaviors \cite{allgeuer2024robots}, and produce illogical communication \cite{kim2024understanding}. These errors undermine the user experience with robotic companions \cite{khan2018reframing} and can even induce anxiety in users \cite{khan2018reframing}. Careful design of robot speech content is critical, as even minimal modifications can alter users' perceptions and engagement with the robot \cite{karli2023if}. While other methods exist for steering LLM outputs (\ie supervised finetuning and in-context learning), prompt engineering is the easiest, most commonly used, and least resource-intensive approach \cite{liu2023pre}, motivating our focus on tools that support prompt engineering.

\subsection{Prompt Engineering}
\label{sec:prompt-engineering}
Prompt engineering is a non-intuitive skill that requires practice and training to perform effectively \cite{oppenlaender2025prompting}. Users frequently struggle with where to begin (“blank page problem”) and tend to rely on opportunistic, ad-hoc trial-and-error prompt tinkering rather than systematic design, often overgeneralizing from a small number of trials and producing weak outcomes \cite{zhang2025chainbuddy, zamfirescu2023johnny}. These challenges are exacerbated by the lack of concrete evidence from interaction logs to ground design decisions and by user feedback that is often overly vague and impressionistic rather than specific and actionable \cite{choi2021protochat}. 

To address these challenges, general-purpose guidance for prompt engineering offers useful heuristics. First, be \emph{clear and specific}: describe desired behaviors affirmatively with measurable targets; avoid vague prohibitions \cite{ibmPrompt, openaiPrompt}. Second, \emph{use positive exemplars}: few-shot examples reliably improve model adherence to target style and structure, as demonstrated for GPT-3 and reinforced in contemporary vendor guides \cite{brown2020language, claudePrompt, googlePrompt}. Third, \emph{reduce imprecision}: supply concrete constraints, inputs, and outputs rather than broad descriptions \cite{claudePrompt}. In addition, prior work have explored ways to support prompt engineering through supporting crowd-testing \cite{choi2021protochat}, LLM-simulated testing \cite{nam2025efficient}, providing a graphical interface for comparison of responses across models and prompt variations \cite{arawjo2024chainforge}, and using a chatbot to understand user requirements and goals, and then generating an editable and interactive starter LLM pipelines (\ie example inputs, prompts to try, queried models, and code evaluator) \cite{zhang2025chainbuddy}. However, to the best of our knowledge, no structured methods or tools exist to support the design of human-robot conversations. In this work, we aim to design and develop a system to facilitate the deliberate design of human-robot conversations.

%% file: system.tex
\begin{table}[ht]
\caption{System Design Objectives (DOs).}
\centering
\begin{tabular}{p{0.95\columnwidth}}
\hline
\textbf{DO1. Provide Interaction Log for Reflection} \\
Conversation designers find it hard to feel confident about their decision due to the lack of concrete evidence (\eg visualization of unusual cases) that they can point to to support and ground their design decisions \cite{choi2021protochat}. Moreover, users often resort to prompt testing with text-based chatbots rather than on-robot testing due to lack of interaction logs to support reflection. Systems should provide conversation logs and help users reflect and ground their design decisions in their interaction experience. \\ \hline

\textbf{DO2. Collect Granular and Grounded Feedback} \\
People often abstract their experiences when providing feedback, resulting in comments that are overly general and impressionistic rather than specific, concrete, and actionable \cite{choi2021protochat}. Systems should support the elicitation and collection of granular, concrete, and actionable feedback grounded in the user's conversational interaction. \\ \hline


\textbf{DO3. Facilitate Translation from Feedback to Prompt} \\
As user feedback tends to be too abstract and users often have trouble articulating their rationales behind their preferences, it is difficult to turn the feedback into concrete design decisions and prompt refinements to guide the robot's behavior \cite{choi2021protochat}. Moreover, prompt engineering is a new type of skill that requires deliberate practice and learning to acquire \cite{oppenlaender2025prompting}, which makes the translation from feedback to effective prompt refinement even more challenging. Therefore, systems should provide support in translating abstract user preferences into prompt refinements. \\ \hline


\textbf{DO4. Overcome ``Blank Page Problem''} \\
People often face uncertainty and difficulty in knowing where and how to begin when prompting LLMs, a challenge known as the ``Blank Page Problem''\cite{arawjo2024chainforge}. Therefore, systems should provide initial guidance and structure to help users get started \cite{gordon2023co, zhang2025chainbuddy}. \\ \hline

\textbf{DO5. Track Prompts \& Design Rationale} \\
Prompt design is an iterative process \cite{arawjo2024chainforge}. Users rely on their own mental model to recall design rationale and refinements across iterations, which can be inaccurate. Systems should enable tracking of prompts and design rationale across iterations. \\ \hline


\end{tabular}
\label{tab:design-objectives}
\end{table}

\section{System: AI-Aided Conversation Engine}

\subsection{System Design}
We conducted a five-stage design process\footnote{All studies reported in this work were approved by our institutional review board (IRB). All participants were compensated at $\$15.00$ per hour.} to identify design objectives and guide system design (see Table \ref{tab:design-objectives}). The process involved reviewing relevant literature and interviewing HRI researchers to validate our understanding of current practice and its challenges, followed by iterative development and testing of prototypes to refine both the design objectives and the system itself. 

\subsubsection{Literature Review}
We reviewed recent HRI literature ($2024--2025$) to examine practices in LLM-powered human-robot conversation design ($n=15$). As discussed in Section \ref{sec:human-robot-conversation-design}, prior work largely relies on prompt engineering when designing human-robot conversations. However, we found no accounts of structured methods or tools for this process. Instead, combined with the research team's own experience and understanding, researchers largely depend on impression-based trial-and-error during human-robot conversation design (see predominant practice in Fig. \ref{fig:teaser}). 

\subsubsection{Interviews}
To confirm our understanding, we conducted three 30-minute semi-structured interviews with HRI researchers (2 female, 1 male) recruited via convenience sampling. Participants included an industry researcher with a PhD in computer science and six years of HRI research experience (remote), a computer science PhD candidate with four years of social robotics research experience (in-person), and a computer science PhD with six years of research experience on conversational agents (in-person). All participants had hands-on experience designing LLM-powered conversational robots. In each session, participants reflected on their current practices and discussed challenges and breakdowns in their workflow. \sally{Through reflective thematic analysis, we identified concepts (subthemes) and then grouped them into high-level themes. Our findings, presented below, align with our literature review:}

\sally{\textit{Iterative prompt refinement workflow.} Prompts were iteratively refined, beginning with quick trials in text-based developer sandboxes (\eg OpenAI Playground), followed by on-robot testing. They ended the design process when they felt the conversation is ``good enough'' based on their observed behavior during on-robot testing.}

\sally{\textit{Use of prompt-engineering practices and LLM assistance.} Participants noted that they try to incorporate recommended practices from prompt engineering guidelines when prompting (\eg including positive examples) and one reported using LLMs to help format prompts and generate additional positive examples.}

\sally{\textit{Misaligned modality for testing.} Text-based chatbot platforms were favored for quick initial trials because they simplify experimentation and provide chat history to facilitate reflection, while robots do not come with such interface (DO1). However, there often exists a disconnect between conversational experiences with text-based chatbots and voice-based robots, which can lead to an increased number of design iterations needed.}

\sally{\textit{Informal evaluation and need for structured feedback.} When participants refined the conversation design, their design decisions were guided by their feelings of ``what worked'' and ``what failed.'' There is a need for methods to elicit more structured, specific, and concrete feedback as opposed to overall impressions and instinctive feelings from testing (DO2).}

\subsubsection{Formative Study}
To explore ways of eliciting more granular and grounded feedback from users (DO2), we recruited two participants (both male, no prior experience designing conversational agents). Each participant completed two five-minute test sessions (one on medical self-diagnosis based on a set of symptoms and one on ranking the importance of a list of items in a sea survival scenario) with a pre-designed social robot. After each conversation, we conducted a semi-structured interview about their experience and suggestions for improvement. 

Initially, participants described their experience in vague terms and only offered high-level impressionistic feedback (\eg ``good,'' ``helpful,'' ``gave good information''). This aligns with prior findings that users tend to provide abstract, non-actionable feedback on chatbot design \cite{choi2021protochat}. To probe further, we printed out conversation transcripts and asked participants to annotate them with different colors and patterns indicating what they liked/disliked or found to be necessary/unnecessary. During the activity, participants began to provide many concrete feedback about the robot's communication style and content verbally. For example, one participant marked ``sorry to hear that you're feeling unwell'' as both ``liked'' and ``necessary,'' explaining that it is important to build rapport with users through expressing sympathy. Transcript annotation enabled participants to move from abstract impressions to granular, concrete feedback (DO2).

We also noticed new challenges. Participants frequently struggled to articulate why they liked or disliked certain components of robot speech (\eg ``I don't really know why. I just liked it.'') and sometimes made contradictory comments. The large amount of detailed but scattered feedback and annotations was also difficult to process and organize (DO3). 

\subsubsection{Initial System Prototype}
We used the insights gathered to develop our initial prototype. We implemented a design interface using React and FastAPI framework, integrated with a custom LLM-powered social robot built in ROS2. \sally{Replicating the platform from prior work \cite{cao2025interruption}, the robot leverages an LLM (gpt-4.1-mini-2025-04-14) to generate contextualized speech, Google text-to-speech voice API (Google en-US-Chirp-HD-F) for audio output and end-of-turn detection. We hand-crafted a bank of facial expressions (happy, satisfied, excited, interested, surprised, and thinking), head positions (left gaze, right gaze, look at screen, left nod, right nod, thinking), and idle behaviors (breathing, blinking). We prompt-engineered a system prompt to 1) format LLM outputs into robot-compatible format, 2) segment robot speech into chucks, and 3) select contextually-appropriate facial expressions and head positions from the bank for each robot speech segment. This system prompt is appended to every robot behavior prompt created in the UI to ensure compatibility with the robot.} We also included the following features: 
\begin{itemize}  [leftmargin=*]
    \item We logged each conversation during testing and displayed the transcript in the UI after each session (DO1). 
    \item \textit{Transcript annotation.} We digitalized the paper-based transcript annotation mechanism. Users can highlight content as ``liked,'' ``disliked,'' ``necessary,'' ``unnecessary,'' ``clear,'' or ``ambiguous''. To complement these tags, users can add free-form comments to the highlighted content, enabling more nuanced feedback (DO2). 
    \item \textit{AI-assisted translation of annotation to feedback.} We incorporated LLMs to facilitate the translation of user annotations into prompt updates (DO3). To increase transparency in design decisions, we separate the refinement process into two components: 1) AI-generated refinement suggestions (gpt-4o-mini) and 2) AI-generated refined prompt (gpt-4o-mini). Our system processes the transcript annotations and comments, summarizes the feedback into prompt refinement suggestions, and translates them into concrete prompt adjustments. Users can review and edit outputs at any step, ensuring both transparency and user control.  
\end{itemize}

\subsubsection{Initial Prototype Testing}
We recruited three participants (2 female, 1 male) to test our high-fidelity prototype in designing a social robot that can act as decision support in ranking items based on survival importance in a sea survival simulation. Two participants had no prior prompt engineering experience, while one had experience designing text-based chatbots for research. The design process took between 30 to 60 minutes. Participants with no prior prompt engineering experience reported difficulty getting started with drafting initial prompts, because they were unsure about what information to include and how to specify and guide the robot behavior (DO4). Moreover, one participant noted that after some refinements, they preferred the robot's earlier behavior but lacked the ability to revert to a previous version (DO5). Based on the insights, we refined our initial prototype:
\begin{itemize} [leftmargin=*]
    \item \textit{AI-assisted initial prompt creation.} We developed an LLM-powered (gpt-4.1-mini) voice-based assistant (Google text-to-speech en-US-Chirp-HD-D voice; distinct from robot voice) that provides structured guidance to help users overcome the ``blank page problem'' (DO4). The assistant co-creates the initial robot behavior prompt by engaging users in dialogue to elicit task goals, define the robot's expected role, and specify stylistic preferences for conversation (see example conversation in Fig. \ref{fig:teaser}). 
    \item \textit{Design Iteration History.} We added prompt, transcript, and respective annotation history to support reflective design (DO5).
    \item We added additional tags (informative, redundant, concise, wordy, on-topic, off-topic, helpful, confusing, polite, rude, and other) to elicit more specific feedback. 
\end{itemize}

\subsubsection{Refined Prototype Testing}
We recruited four participants (2 female, 2 male), none of whom had prior experience designing conversational agents, to test our high-fidelity prototype on the same sea survival decision-support robot design task. The design process lasted between 25 and 70 minutes. Overall, participants found the tool to be ``easy to use.'' When asked to rate their satisfaction with their final prompt design on a scale of 0 to 100, three participants gave a score of 100, while one participant, who only went through one design cycle (initial prompt $\to$ test $\to$ refine), gave a score of 85. Based on our observations of the user's design process and a review of the prompts created, we made the following adjustments to our prototype: 
\begin{itemize}  [leftmargin=*]
    \item We incorporated relevant recommended prompt engineering practices into both the initial and refined prompt generating LLM to improve the LLM's ability to translate feedback into an effective prompt (DO3).
    \item The suggested improvements were too long, and participants did not seem to read through the suggestions carefully from the screen recording. To improve readability, we restructured the suggestions into four concise, bulleted lists each focused on ``Essential Behaviors to Maintain'', ``Behaviors to Reduce or Avoid'', ``Positive Engagement Cues'', and ``Recommended Adjustments.'' 
    \item We removed ``polite,'' ``rude'', and ``other'' from the set of tags as they were not used.
\end{itemize}

Lastly, we conducted a pilot study ($n=3$) to verify changes made to the prototype and made minor adjustments to the UI and instructions to improve the usability of the interface based on feedback.

\subsection{Final System Overview}
\begin{figure}
\centering
\includegraphics[width=\columnwidth]{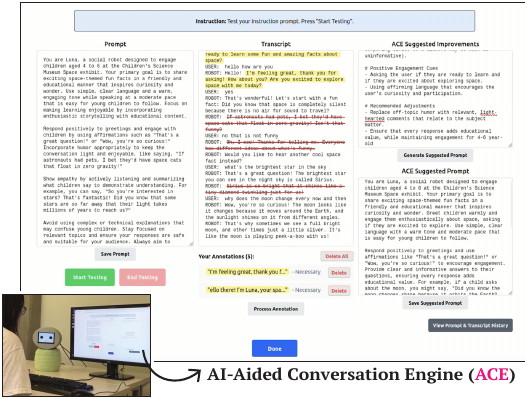}
\caption{Main Design Page of ACE Interface. See supplementary materials for an enlarged view of the interface.}
\Description{An image of the interface with prompt on the left, transcript in the middle, AI suggested Improvement panel and ACE Suggested Prompt panel on the right. In the corner, there is an image from the back view of a participant working on a design conversation for a social robot that is sitting on the table.}
\label{fig:inteface}
\end{figure}

Based on design objectives identified from the five-stage design process, we developed AI-Aided Conversation Engine (ACE) to support deliberate design of human-robot conversation (Fig. \ref{fig:inteface}). Our system distinguishes from predominant practice in five ways as illustrated in Fig \ref{fig:teaser}: We integrate the social robot into the design interface to allow to easy testing and visualizes the conversation log for reflection (DO1); we enable the collection of granular and grounded rather than impression-based feedback to support reflection (DO2); we use LLMs to facilitate the translation of collected feedback into prompt refinements (DO3); we use voice-based LLM agent to scaffold and support overcoming ``blank page problem'' (DO4); and we enable the tracking of prompts and design rationales across iterations to facilitate reflective design (DO5).

%% file: study1.tex
\begin{figure*}
\centering
\includegraphics[width=0.9\textwidth]{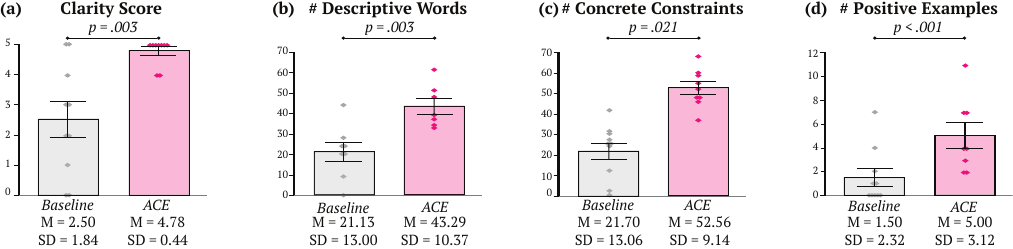}
\caption{Effects of using Baseline and ACE in human-robot conversation design on prompt clarity (a), prompt specificity (b, c, d). Error bars shown represent standard error.}
\label{fig:results1}
\Description{(a) Bar plot of prompt clarity score. ACE had on average prompt clarity score of 4.78, SD=0.44. Baseline had on average prompt clarity score of 2.50, SD=1.84. (b) Bar plot of the number of descriptive words. ACE had on average 43.29 descriptive words, SD=10.37. Baseline had on average 21.13 descriptive words, SD=13.00. (c) Bar plot of number of concrete constraints. ACE had on average 52.56 concrete constraints, SD=9.14. Baseline had on average 21.70 concrete constraints, SD=13.06. (d) Bar plot of the number of positive examples. ACE had on average 5 positive examples, SD=3.12. Baseline had on average 1.50 positive examples, SD=2.32.}
\end{figure*}

\begin{figure}
\centering
\includegraphics[width=0.9\columnwidth]{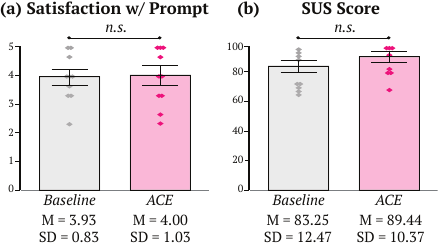}
\caption{Effects of using Baseline and ACE on (a) participants' perceived satisfaction with the final prompt (b) SUS score. Error bars shown represent standard error.}
\Description{(a) Bar plot of satisfaction with prompt. ACE had on average score for satisfaction with prompt of 4.00, SD=1.03. Baseline had on average score for satisfaction with prompt of 3.93, SD=0.83. (b) Bar plot of SUS score. ACE had on average SUS score of 89.44, SD=10.37. Baseline had on average SUS score of 83.25, SD=12.47.}
\label{fig:results1-2}
\end{figure}

\section{Study 1: Evaluation from Designer Perspective}
We conducted a user study to evaluate the usability and effectiveness of our system in facilitating human-robot conversation design. 

\subsection{Comparison to Baseline System}
As a baseline to compare our proposed system against, we implemented an interface (see supplemental materials) similar to commercial prompt engineering or chatbot prototyping interfaces, \eg OpenAI Platform Playground or Claude Console. We add instructions to facilitate designers, advising them to ``consider describing the task, robot's expected role, and any other desired communication style and behavior guidelines for the robot.'' In addition, we added a history feature to allow participants to see their previous prompt and respective transcripts from the test interactions. Lastly, we integrated the interface with the same basic LLM-powered social robot as ACE to allow for fluid testing. In our study, the design tool (Baseline vs. ACE) was a between-subjects factor. 


\subsection{Experimental Task}
Participants were told they had been hired as a robot conversation designer for a children's science museum. The robot's role was to share space-themed fun facts with children aged 4--6 visiting the museum's space exhibit. Participants were tasked with designing a conversational interaction that is engaging, fun, and age-appropriate for young children. The interaction should help children become more curious and excited about the exhibit. We chose this task since it is distinctive from the tasks from the formative study and since designing engaging content for young children requires creativity and is easy to relate to. Participants had unlimited time to design their robot and were told that their robot designs would later be evaluated by users, and the designer for the best rated human-robot conversation would receive $\$30$ bonus reward. 

\subsection{Study Procedure}
After providing consent, participants were introduced to their role as a "robot conversation designer". They watched a 1-minute video on a simple sample conversation with the social robot on "places to visit in Taipei". Then, the user was provided with design instructions that included a walk-through of the design interface (Baseline or ACE). The design process began after they completed a demographic questionnaire. After the design process, the participants completed a post-study questionnaire on their experience with the design interface. The study concluded with a semi-structured interview aimed to understand the participants' overall experience using the design tool and any feedback they may have. Participants who have had prompt engineering experience were asked to compare their experience designing the conversational robot using the interface with their prior experiences. Participant interactions were screen, video, and audio recorded.

\subsection{Participants}
We recruited 20 participants (10 baseline; 10 experimental) through convenience sampling from the local community, using physical flyers and electronic posts to community newsletters and mailing lists. One participant was excluded from our analysis due to a lack of engagement and contradictory responses across measures and the interview. Of the 19 participants (10 female, 9 male), their age ranged from 18 to 27 ($M=22.26, SD=3.05$). Nine of the participants' most recent field of study is computer science (5 baseline; 4 experimental). Participants were moderately experienced with AI ($M=3.11, SD=0.81$, 5-point scale with 1 being not at all familiar and 5 being extremely familiar). However, most participants have never or have only tried prompt engineering a few times ($M=2.26, SD=1.10$, 5-point scale with 1 being never used and 5 very frequently used) and were not confident in their ability in prompt engineering ($M=2.53, SD=1.07$, 5-point scale with 1 being very not confident and 5 being very confident). The study took roughly $60$ minutes. 

\subsection{Measures}
We evaluated the systems on the quality of the generated behavior prompts, the design process, and the usability of the system. 


\subsubsection{Behavior Prompt Quality} 
\sally{We evaluated prompt quality following prompt engineering guidelines along two dimensions: clarity and specificity. \textit{Clarity score} ($0$--$5$) was coded by awarding one point each for describing the robot's task, relevant task context, robot's role, audience, and desired output style. Specificity was assessed by counting the \textit{number of descriptive words} (adjectives and adverbs), \textit{number of constraints}, and \textit{number of examples} in the prompt. One coder independently coded all final prompts created by participants, while a secondary coder independently coded $20\%$ of the data. Intercoder reliability was assessed using the intraclass correlation coefficient (ICC); ICC above 0.90 indicates excellent reliability. All three metrics had excellent coded reliability (clarity score: $0.98$, number of constraints: $0.99$, number of examples: $0.97$)\cite{koo2016guideline}. In addition, we measured \textit{user satisfaction with final prompt} using a three-item 5-point Likert scale questionnaire\footnotemark[1] ($0$--$5$), capturing designer satisfaction and confidence (Cronbach's $\alpha = 0.91$).}

\subsubsection{Usability}
We assessed usability using the System Usability Scale (SUS), a 10-item, five-point Likert scale questionnaire \cite{brooke1996sus}. SUS scores above $70$ ($0$--$100$) indicate good system usability \cite{bangor2009determining}.


\subsection{Results}
In the analyses reported below, we performed Welch’s t-tests assuming unequal variances to examine the effect of using ACE compared to the baseline tool (Baseline). Baseline and ACE participants were comparable in terms of usage. Participants began the design process by spending an average of $5.11$ minutes ($SD=2.80$) preparing the initial prompt with ACE and $4.29$ minutes ($SD=3.65$) with Baseline. Then, on average, participants with ACE engaged in $4.29$ ($SD=3.65$) design cycles, whereas those with Baseline completed $5$ design cycles ($SD=2.80$). In both conditions, during each design cycle, participants tested their prompt and refined it based on what they liked and disliked during the interaction. Participants with ACE ($M=3.66, SD=1.72$) spent on average more time per design cycle on prompt refinement than those with Baseline ($M=2.10, SD=1.21$). Overall, participants using ACE spent an average of $39.73$ minutes ($SD=14.87$) designing the robot behavior, while those using Baseline spent $40.78$ minutes ($SD=13.06$). 







\subsubsection{Participants with ACE crafted prompts with higher clarity and specificity than Baseline} \mbox{}\\
Final prompts crafted with ACE had significantly higher clarity than those with Baseline, $t(10.14)=-3.79, p=.003$ (Fig. \ref{fig:results1}~a). Moreover, final prompts crafted using ACE had a significantly more descriptive words ($t(12.92)=-3.67, p=.003$; Fig. \ref{fig:results1}~b), concrete constraints ($t(12.65)=-2.64, p=.021$; Fig. \ref{fig:results1}~c), and positive examples ($t(16.10)=-6.01, p<.001$; Fig. \ref{fig:results1}~d) than final prompts crafted using Baseline. However, we did not find any significant differences in user satisfaction with the final prompt created using ACE versus baseline, $t(15.41)=0.15, p=.879$ (Fig. \ref{fig:results1-2}~a). 





\subsubsection{Participants found ACE to be easy to use} \mbox{}\\
During post-study interview, five out of ten participants described Baseline as either easy to use ($n=3$), simple ($n=1$), or intuitive ($n=1$), and seven out of nine ACE participants described it as either easy to use ($n=4$) or intuitive ($n=4$). The average SUS score for Baseline was $83.25$, and $89.44$ for ACE, both of which are above the $80$ threshold for excellent (acceptable) usability. However, we did not find significant differences in the SUS score between ACE and Baseline, $t(16.91)=1.18, p=.254$ (Fig. \ref{fig:results1-2}~b).

%% file: study2.tex
\begin{figure}
\centering
\includegraphics[width=\columnwidth]{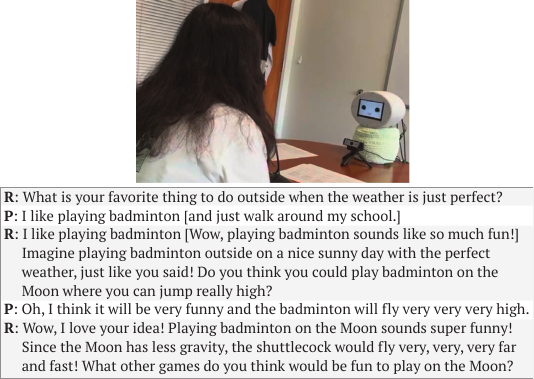}
\caption{Sample dialogue from Study 2 with robot using ACE prompt. Text in brackets indicates overlapping speech.}
\Description{Image from the back view of a participant talking to a social robot. R: What is your favorite thing to do outside when the weather is just perfect?
P: I like playing badminton [and just walk around my school.] R: I like playing badminton [Wow, playing badminton sounds like so much fun!] Imagine playing badminton outside on a nice sunny day with the perfect weather, just like you said! Do you think you could play badminton on the Moon where you can jump really high? P: Oh, I think it will be very funny and the badminton will fly very very very high. R: Wow, I love your idea! Playing badminton on the Moon sounds super funny! Since the Moon has less gravity, the shuttlecock would fly very, very, very far and fast! What other games do you think would be fun to play on the Moon?}
\label{fig:sample-study2}
\end{figure}

\section{Study 2: Evaluation from End User Perspective}
While participants who acted as robot conversation designers found ACE helpful and easy to use, we also wanted to investigate whether final prompts with higher clarity and specificity would lead to differences in interaction quality. To this end, we conducted a second user study to evaluate the prompts created in Study 1. 


\subsection{Experimental Task}
\sally{Participants were told they had been hired by a children’s science museum to design a new space-themed exhibit. The exhibit will feature a robot whose role is ``to share space-themed fun facts with children aged 4–6.'' The interaction is expected to be ``engaging, fun, and age-appropriate for young children.'' A robot conversational designer has designed the robot’s speech content and created two alternative designs. Participants were tasked to test the two designs, evaluate them on engagement, fun, and age appropriateness, and decide which robot design to use for the exhibit. We used all robot behavior prompts created by participants in Study 1, presented in random order. Each prompt was used twice, with two additional randomly selected ACE-generated prompts repeated once more to balance conditions. Hence, each participant engaged with the robot once with a randomly selected ACE-generated prompt and once with a randomly selected Baseline prompt.}

\subsection{Study Procedure}
Upon providing consent, participants were introduced to the task and provided a rubric outlining criteria for engagement, fun, and age appropriateness of the conversation. Then, they watched a one-minute video demonstrating a sample conversation with the social robot on "places to visit in Taipei". Afterwards, participants engaged with each of the two robot designs for five minutes (see sample interaction in Fig. \ref{fig:sample-study2}). After each session, participants completed the rubric and indicated whether they preferred the first or second robot for inclusion in the exhibit. The study concluded with a semi-structured interview that probed whether participants noticed differences between the two robot designs and what they entailed.

\subsection{Measures}
After testing each robot design, participants rated \sally{\textit{task success} (whether the robot told a fun fact; true/false)} and \textit{goodness of interaction} ($0$--$5$) using a 13-item questionnaire (Cronbach's $\alpha = 0.92$) assessing engagement, fun, and task suitability. At the end, participants selected their preferred design for the museum exhibit.

\subsection{Participants}
We recruited 20 new participants (11 female, 9 male), aged $20$ to $31$ ($M=23.9, SD=2.95$), through convenience sampling using physical flyers and electronic posts to community newsletters and mailing lists. Most participants were familiar with AI ($M=3.95, SD=0.94$, 5-point Likert scale with 1 being not at all familiar and 5 being extremely familiar). However, $82.60\%$ of participants indicated that they use voice-based AI tools less than a few times a month. The study took roughly $30$ minutes. 

\begin{figure}
\centering
\includegraphics[width=0.9\columnwidth]{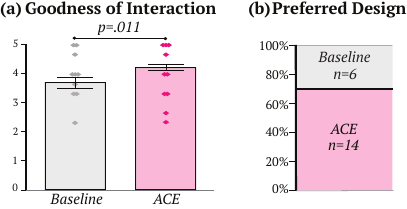}
\caption{Effects of using Baseline and ACE prompts in human-robot conversation on a) goodness of interaction rating (bars represent standard error) and b) preferred design.}
\Description{(a) Bar plot of goodness of interaction. ACE has a significantly higher goodness of interaction than Baseline. (b) stacked plot of preferred design between Baseline and ACE. There are 6 participants who preferred Baseline and 14 participants who preferred ACE.}
\label{fig:results2}
\end{figure}

\subsection{Results}
\sally{Participants rated the robot as successfully completing the task for all but one baseline interaction. Moreover, \textbf{robots with ACE prompts received higher goodness of interaction rating}. A one-tailed pooled t-test revealed that participants rated interaction with robots with ACE prompts ($M=4.19, SD=0.49$) to be significantly better than robots with Baseline prompts ($M=3.67, SD=0.83$), $t(38)=-2.40, p=.011$ (Fig. \ref{fig:results2}~a). $14$ out of $20$ ($70\%$) of participants selected the design generated by ACE in the end (Fig. \ref{fig:results2}~b). One-tailed binomial test indicated that this was not significantly different from the expected proportion of $50\%$, $p=0.058$.}


%% file: discussion.tex
\section{\sally{Discussion and Conclusion}}
This work presents ACE, an open-source tool\footnotemark[1] designed to support the deliberate design of LLM-powered human-robot conversations. In a designer-focused evaluation study, although Baseline and ACE participants spent a similar amount of time on average in the design process, ACE helped participants create robot behavior prompts with greater clarity and specificity. In an end-user-focused evaluation study, prompts crafted with ACE yielded high-quality interactions in the target context. Overall, ACE lowers barriers and is a step toward enabling the deliberate design of more effective, engaging, and appropriate human-robot conversations.

\subsection{Supporting Users in Human-Robot Conversation Design}
In Study~1 interviews, participants highlighted the usefulness of using highlights to provide feedback (DO2) and in leveraging generative AI to translate the often still abstract feedback in annotations into concrete, actionable prompt improvements (DO3). A Baseline participant with no prior prompt engineering experience noted that \pquotes{``knowing how do I adjust the prompt to change the conversation was a challenge''}. In contrast, an ACE participant with no prior prompt engineering experience noted that \pquotes{``I feel like it was easy for me to prompt the AI to be like `oh, I didn't liked this,' 'oh, I liked this', and like 'why,' and stuff like that, and it took those into account''} and \pquotes{``it was accurate in generating prompt improvements based on the annotations''}. This feature was not only appreciated by participants new to prompt engineering, but also liked by experienced prompt engineers. Both ACE participants who frequently engaged in prompt engineer described their experience prompting engineering with ACE to be \pquotes{``better''} than their prior prompt engineering experience. Both of them found that \pquotes{`` highlighting was really helpful''} and noted that \pquotes{``ACE was really good at explaining things, understanding things, and preparing an actually really good prompt.''}

Moreover, participants noted that ACE, integrated with generative AI, not only \pquotes{``cuts out a lot of manual work''} (\eg \pquotes{``formatting things during prompt generation''}), but also provided inspiration and augmented creativity during initial prompt generation and reflective refinement. Participants described ACE as \pquotes{``good at coming up with suggestions''} during initial prompt creation. Rather than asking a fixed set of pre-defined questions, ACE requested additional details when needed and offered ideas to help users brainstorm (see sample conversation in Fig. \ref{fig:teaser}). One participant \pquotes{``really appreciated that when [they were] annotating the response of [the robot], it took [their] response into account and added things that [they] wouldn't have thought of.''} Because user intentions behind the annotations can be vague or abstract, ACE was designed to infer patterns from annotated transcripts and translate them into clear behavioral adjustment recommendations (DO3). While the suggested improvements did not always exactly match what users had in mind, such mismatch can spark new ideas and facilitate reflection \cite{antony2025id}. 

Through these features, ACE effectively supported participants with and without prior prompt engineering experience in crafting context-specific human-robot conversations. In particular, five ACE participants had no prior prompt engineering experience in the designer-focused study. Robots using prompts designed by them were, on average, rated $4.13$ out of $5$ for goodness of interaction in the end-user-focused evaluation study ($n=12$), indicating that the conversational interaction was suitable for context.


\subsection{Tools for Holistic Design of Human-Robot Conversational Experience}
ACE supported users in designing a single LLM prompt for the context that only takes into account the transcribed user speech and conversation history during behavior generation. However, users expect more when designing robot conversations. For example, participants wanted finer control over robot behavior in their prompt (\eg \prompts{``If you hear a response containing ``yes'' or ``yeah'' continue sharing space facts. If you hear a response containing ``no'' share a fun little goodbye encouraging the user to come back if they want to learn more.''}). Such branching would be better implemented via an intent classifier LLM whose output routes to multiple specialized LLMs (continue vs. conclude) \cite{cao2025interruption}. Moreover, some participants tried to control the robot's speaking rate in their prompt (\eg \pquotes{``Speak with a natural, conversational pacing that is neither too fast nor too slow.''}). However, the text-to-speech settings were fixed and did not adapt to the prompt. Other participants also tried to incorporate timing (\eg \pquotes{``provide a new fun fact about anything in the entire space for every 5 seconds of silence.''}) and user mental state (\eg confusion, disengagement, uncertainty, frustration) into their prompt (\eg \pquotes{``Be attentive to the children's emotional state. If they express frustration or disengagement, gently shift to lighter topics or fun facts to re-engage them, such as sharing a whimsical fact about a funny space phenomenon.''}) as triggers for robot behavior. Future work should explore ways to enable users to incorporate user mental state \cite{stiber2024uh, grafsgaard2013automatically, scherf2024you} and interaction state estimations \cite{kontogiorgos2021systematic, cao2025err, spitale2025vita} using multi-modal inputs to support the design of more adaptive and tailored human-robot conversations. Furthermore, participants expressed interest in designing the voice and robot non-verbal behaviors (facial expressions, body movement, and looks of the robot). \sally{In summary, while robot speech content is the basis for most behavior generation pipelines, as non-verbal behaviors are typically generated contingent upon speech (e.g., co-gestures), future work should investigate design tools for a more holistic design of human-robot conversational experiences by considering multi-modal inputs and outputs, which are essential components of embodied interaction.} 

\subsection{Limitations}
Our evaluation was limited \sally{by a small sample size, convenience sampling, and} single in-lab interactions. Future work should assess whether ACE prompts can sustain engagement in longer-term, multi-session interactions in the wild \sally{engaging domain experts as designers} \cite{jain2020modeling}. Moreover, while ACE effectively supported novice prompt engineers, future work should explore whether ACE can support actual stakeholders while designing human-robot conversations for their specific application domain \cite{mahmood2025voice, antony2023co}. Additionally, while ACE helped users create prompts with greater clarity and specificity, one participant observed that over time, as their prompts became increasingly detailed and contained more positive examples, the LLM output converged to their expectation, but felt less diverse and creative. This suggests that there may be a need to balance the precision and appropriateness of design with variance, diversity, and creativity. Future work is needed to explore this potential trade-off and prevent ``over-engineering'' prompts. Finally, while we focused on designing and evaluating ACE to support the design of human-robot conversations, ACE can be used to support the design of human-agent conversations in general.